\documentclass[
]{ceurart}

\usepackage{listings}
\usepackage{subcaption}
\usepackage{caption}
\usepackage{xspace}
\usepackage[capitalize, nameinlink]{cleveref}

\begin{document}

\copyrightyear{2024}
\copyrightclause{Copyright for this paper by its authors.
  Use permitted under Creative Commons License Attribution 4.0
  International (CC BY 4.0).}

\newcommand{\myparagraph}[1]{\vspace{0.25em}\noindent \textbf{#1.}}
\newcommand{\minio}{MinIO\xspace}

\conference{Workshop on Knowledge Management, Trustworthiness, Interpretability in AI and Beyond - CIKM 2023}

\title{A Flexible and Scalable Approach for Collecting Wildlife Advertisements on the Web}

\tnotemark[1]

\author[1]{Juliana Barbosa}[%
email=juliana.barbosa@nyu.edu,
]
\address[1]{New York University}

\author[2]{Sunandan Chakraborty}[%
email=sunchak@iu.edu,
]
\address[2]{Indiana University}

\author[1]{Juliana Freire}[%
orcid=0000-0003-3915-7075,
email=juliana.freire@nyu.edu,
]


\begin{abstract}
Wildlife traffickers are increasingly carrying out their activities in cyberspace. 
As they advertise and sell wildlife products in online marketplaces,  they leave digital traces of their activity. This creates a new opportunity: by analyzing these traces, we can obtain insights into how trafficking networks work as well as how they can be disrupted.
However, collecting such information is difficult. Online marketplaces sell a very large number of products and identifying ads that actually involve wildlife is a complex task that is hard to automate. Furthermore, given that the volume of data is staggering, we need scalable mechanisms to acquire, filter, and store the ads, as well as to make them available for analysis. 
In this paper, we present a new approach to collect wildlife trafficking data at scale. We propose a data collection pipeline that combines scoped crawlers for data discovery and acquisition with foundational models and machine learning classifiers to identify relevant ads. We describe a dataset we created using this pipeline which is, to the best of our knowledge, the largest of its kind: it contains almost a million ads obtained from 41 marketplaces, covering 235 species and  20 languages.
The source code is publicly available at \url{https://github.com/VIDA-NYU/wildlife_pipeline}.
\end{abstract}


\maketitle

\begin{figure}
  \centering
  \includegraphics[width=\textwidth]{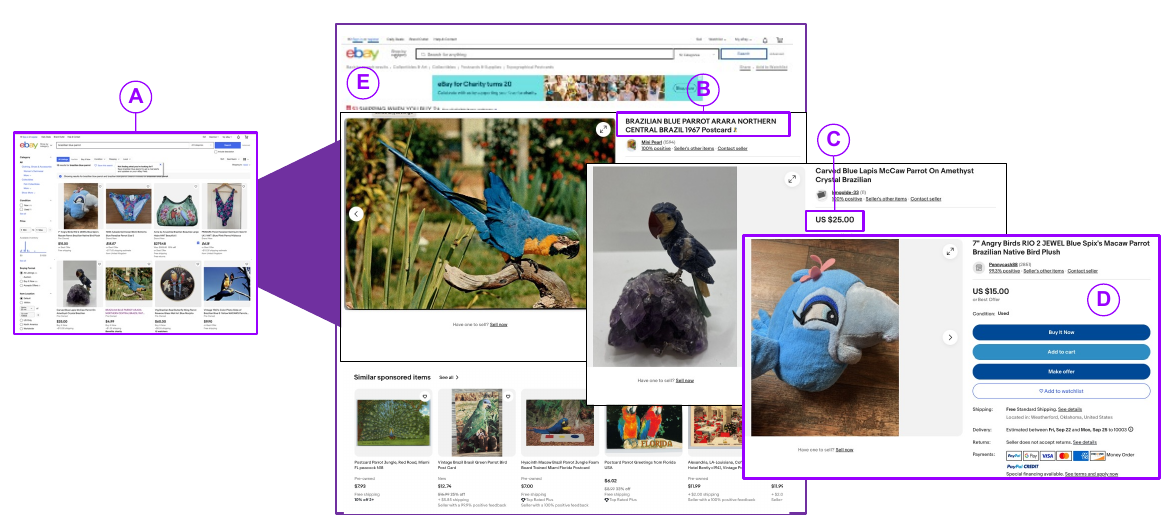}
  \vspace{-.3cm}
  \caption{We describe a pipeline that automatically collects a dataset of wildlife advertisements (ads) from e-commerce websites. Issuing a 
  query (A) ``Brazilian blue parrot", we can find and follow links to several product pages (E). For each product (D), we extract product attributes such as (B) product title and (C) price. A challenge in constructing this dataset is how to distinguish ads for wildlife products from other types of products, such as postcards and toys.}
  \label{fig:teaser}
\end{figure}

\section{Introduction}

Wildlife trafficking is one of the most common illicit activities, with human, social, and economic consequences \cite{magliocca2021comparative}. It not only exacts a considerable toll on society through increased criminality \cite{anagnostou2022illegal} and environmental devastation \cite{maxwell2016biodiversity}, but may also expose us to unforeseen health and bio-safety hazards \cite{aguirre2021opportunities,keskin2022quantitative}.

Globalization has had a significant impact on the wildlife trade, leading to a notable expansion from traditional markets to online platforms. Several factors contribute to this transformation, such as increased connectivity, more efficient supply chains, and the rise of e-commerce platforms \cite{siriwat2020wildlife}. 
While this creates challenges, it also opens new opportunities. As criminals use technology, they leave traces of their activity on the web. 
These traces have been used in studies
that have significantly improved our understanding of online wildlife trafficking. But these studies are often limited to specific species~\cite{harrington2019popularity, martin2018trade, gomez2019bearly, venturini2020disguising, roberts2022systematic}, regions~\cite{vietnamese2016rapid, roberts2022systematic, stoner2014tigers} or specific sites. To better understand the traffic of wildlife, it is crucial to have large-scale data resources~\cite{haas2015federated, charity2020wildlife} that have a broader coverage of species, regions, and platforms.

\myparagraph{Collecting Wildlife Ads at Scale: Opportunities and Challenges} Online marketplaces are a rich source for obtaining traces of wildlife trafficking \cite{hastie2014wanted}. 
Ads obtained from these sites make it possible for researchers to answer important questions regarding the general dynamics of the online trade, e.g.,  its volume, species targeted, their origin parts sold, and asking prices\cite{stoner2014tigers}. However, this opportunity comes with several challenges. 
Finding and collecting data to support uncovering illicit activities is difficult~\cite{keskin2022quantitative}.
Online marketplaces sell many different types of products, hence a keyword-based search in such sites retrieves products belonging to a variety of categories. 
\autoref{fig:teaser} we can see 3 possible results that can be found on a search for ``Brazilian blue parrot": a postcard, a parrot carved on crystal, and a stuffed parrot -- no real animals are returned.
Identifying ads that actually include animals and animal parts for sale is important to streamline the data collection, i.e., to reduce the retrieval of irrelevant pages, and for the downstream data analysis.

Machine learning classifiers can be trained to perform this identification. But the process of training a new model is complex and costly due to the scarcity of labeled data. Moreover, traffickers deliberately hide their actions, making the acquisition of suitable training data, characterized by its authenticity and specificity, even harder to obtain.
For these data to be useful for analysis, we must first extract structured information about the products (the attributes in an ad, such as title, price) from the unstructured product pages. Since the structure and format of different sites vary widely with respect to layout, content structure, and presentation, it is challenging to extract data consistently \cite{wang2022webformer, lin2020freedom, dong2020multi}.
Often, extraction scripts (scrapers) must be hand-crafted for specific websites.
Embedded metadata from HTML markup is another source of structured data for products. But just like the HTML descriptions, it can come in different formats and not all sites publish the information. 
 
This problem is compounded due to the fact the sites are dynamic and often change how their pages (and metadata) are structured \cite{santos2016first}. Consequently,
scaling up the data collection process to find  and cover a large number of sites can be time-consuming, both to create and continuously update the scrapers \cite{pham2018learning, pham2019bootstrapping}.

\myparagraph{Our Contribution} We take a first step towards enabling the large-scale collection of online wildlife ads. We have designed a flexible pipeline to collect product pages that are published on different sites and extract information that is useful for analysis and exploration of wildlife trafficking, such as price, images, and sellers. 
As we discuss in Section~\ref{sec:pipeline}, this pipeline is scalable and flexible: it can be customized for different types of collections (e.g., the set of species of interest, the platforms to be crawled);  integrate a wide range of models to identify relevant products and perform extraction; and store data in easily accessible cloud storage platforms such as s3.  
To demonstrate the effectiveness of the pipeline, we describe a preliminary dataset we derived using web pages collected over 34 days which contains almost a million ads from 41 marketplaces, covering 235 species and 20 different languages.

\myparagraph{Related Work}
Our work is related to the field of data collection, focused on the domain of illicit wildlife trade. Keskin et al.  \cite{keskin2022quantitative} offers a comprehensive perspective on the illicit wildlife trade, shedding light on the critical challenges encountered in this domain. One of the most significant challenges highlighted is the scarcity of data, exacerbated by the fact that the available data tends to be biased toward specific regions and particular species. For instance, Cardoso et al.~\cite{cardoso2023detecting} trained a neural network to identify pangolins. Kulkarni and Di Minin \cite{kulkarni2023towards} highlight the increasing efforts towards wildlife conservation using machine learning, but confirm the challenges of finding good quality labeled data.

Given that e-commerce has become a new point for wildlife trade, some studies analyzing web-crawled data \cite{xu2020illegal} are emerging. Many of the existing studies have used manual searches to gather evidence of the online wildlife trade, particularly products from endangered species~\cite{ifaw2008killing,traffic2019}. This labor-intensive approach though accurate has limitations, in terms of being slow and hard to scale. Automated computational methods have emerged as a promising solution, as they can continuously monitor a wide range of species across the digital landscape without heavy reliance on human resources. However, existing studies focusing on automated detection have been used to identify very specific endangered species like cacti~\cite{sajeva2013regulating}, elephant ivory~\cite{hernandez2015automatic}, and orchids~\cite{hinsley2016estimating}.

Our research aims to address the data gap by enabling the creation of diverse and extensive datasets that include a wide range of animal species and web-market locations. The dataset we describe is, to the best of our knowledge, the first to cover a large variety of endangered species being advertised in multiple countries.

\vspace{-.3cm}
\section{Data collection Pipeline
\label{sec:pipeline}}
\vspace{-.15cm}

\begin{figure*}[t]
\begin{center}
\includegraphics[width=\textwidth]{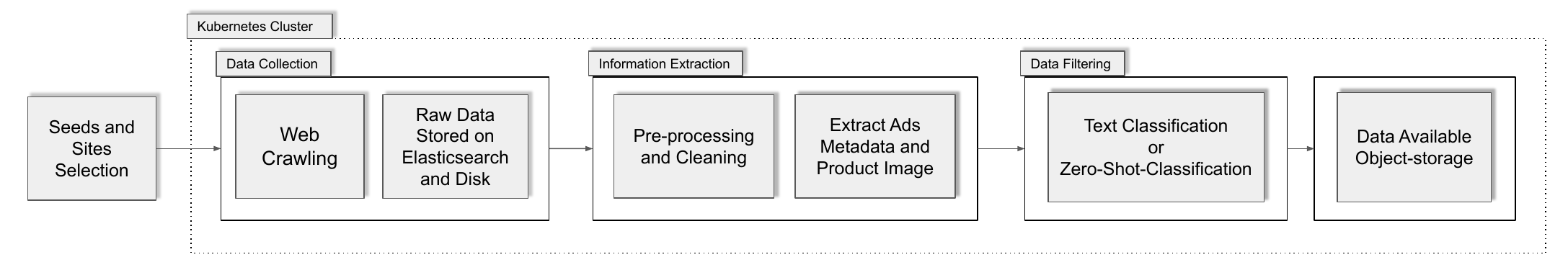}
\vspace{-.6cm}
\caption{Data Collection Pipeline.} 
\vspace{-.6cm}
\label{fig:data-collection-pipeline}
\end{center}
\end{figure*}

\autoref{fig:data-collection-pipeline}
provides an overview of the key components of the data-collection pipeline. We describe them in detail below.

\myparagraph{Seeds and Site Selection}
To collect pages using a web crawler, we need to provide as input a set of \emph{seed} URLs that serve as entry points for the crawl.
The crawler follows each seed, downloads the corresponding web pages, and recursively follows links extracted from these pages. 

To obtain wildlife-related ads, we generate the seeds dynamically, based on the underlying URLs that e-commerce websites use for their search ``forms", combined with a list of names of endangered animals as search queries. Since different websites have distinct search query patterns, thus we need to compile a list of patterns for the selected sites. For example, if we want to search for ads on \texttt{ebay.com} using the query \texttt{KEYWORD}, the pattern of URL for the search form is:
%
{\color{MidnightBlue}\url{/sch/i.html_from=R40&_nkw=}\color{Blue}\textbf{\texttt{KEYWORD}}\color{MidnightBlue}\url{&_sacat}}.
%
Then, we can substitute the query \texttt{KEYWORD} with the names in the list of endangered animals.

To construct our dataset, in collaboration with domain experts, we created a list of animals provided by the Convention on International Trade in Endangered Species of Wild Fauna and Flora (CITES) that includes endangered species (\textit{CITES Appendix I}) -- 
species that are threatened with extinction and are therefore provided with the highest level of protection. 

We generated patterns for a curated list of 49
distinct e-commerce websites, which include 20 eBay platforms operating in various countries.
The initial list of websites was selected from the results emerging from web searches using keywords related to the species and other wildlife products. For example, we obtain the top sites resulting from search queries (e.g.,``tiger taxidermy''), and sites that appear in multiple results were compiled to form the initial set. 

We use the species name and their respective English names as keywords to generate the seeds. Each of these animals can have one or more English names, leading to a total of 1017 keyword queries. The final list contains 49,833 seeds that include one unique URL for each keyword in each domain (see \Cref{fig:teaser}(A)). In Section \ref{sec:dataset} we describe in detail the dataset collected from these seeds.
Note that our approach is general and other lists of animals and websites can be used, depending on the goals of the data collection.

\begin{table}[t]
    \small
    \centering
    \caption{Data overview: attributes and their descriptions, and the number of records that contain the attributes.}
    \begin{tabular}{c c c}
        \toprule
        \textbf{Attributes} & \textbf{Description} & \textbf{\# of records}  \\
        \midrule
        \texttt{url} & The ad URL & 954,684 \\
        \texttt{title} & Product Advisement title & 946,732 \\
        \texttt{text} & The page text & 954,684 \\
        \texttt{product} & Name of the product & 954,684 \\
        \texttt{description} & Description of the product & 805,449 \\
        \texttt{domain} & Website where the product is posted & 954,684 \\
        \texttt{image }& URL of the image & 787,185 \\
        \texttt{retrieved }& time when the page was downloaded & 954,684 \\
        \texttt{category} & The category listed for that product & 25,038 \\
        \texttt{production date} & Production date of the product & 5,786 \\
        \texttt{price} & Price of the product & 682,652 \\
        \texttt{currency} & Currency of the price & 679,717 \\
        \texttt{seller} & Seller name & 8,910 \\
        \texttt{seller\_type} & the category the seller is listed & 27,483 \\
        \texttt{location} & Location of product & 25,150 \\
        \texttt{zero\_shot\_label} & zero shot classifier results & 954,684 \\
        \texttt{zero\_shot\_prob} & zero shot label probability & 954,684 \\
        \texttt{id} & UUID used as filename for images & 954,684 \\
        \bottomrule
    \end{tabular}
    \label{table: dataset-description}
\end{table}

\myparagraph{Data Collection}
We use the open-source ACHE crawler to perform a \emph{scoped crawl} \cite{barbosa2007adaptive, ache@github} starting with the 49,833 seeds described above.
ACHE downloads all pages from the seed URLs and extracts the links from the pages. Instead of following all links, ACHE scopes its search and only follows links in the domains associated with the seed URLs. 

\myparagraph{Information Extraction} The crawler retrieves a set of pages that includes ads for individual products. As illustrated in Figure~\ref{fig:teaser}(E), these pages have a lot of information that is not pertinent to the actual product, including eBay ads and sponsored items. Therefore, we need to identify and extract the information associated with the product so that we can produce, for each product, a record containing its relevant attributes, e.g., \textit{Price, Seller, Product type, Description, and Product image}.
But doing so is challenging due to the 
diversity of content and structure used by different sites. 
We implemented a set of strategies that we combined to address this challenge.

\noindent
\textit{Extract page content.}  We use BeautifulSoup \cite{beautifulsoup} to parse the HTML content on the page and extract the page title and text content. 

\noindent
\textit{Extract product attributes.} We scrape the HTML page to extract specific attributes such as price and seller. With MLscraper \cite{mlscraper@github}, we can automatically perform this extraction by just providing the tool with a few examples of the desired output based on the information available on a sample HTML page. 

\noindent
\textit{Extract product metadata.} Some pages have embedded metadata that contains information about the product. To obtain this information, we use Extruct, \cite{extruct@github}, a library designed for extracting embedded metadata from HTML markup. 

\myparagraph{Data Filtering}
\autoref{table: dataset-description} shows a summary of the data derived by the extraction step. While there are almost 1 million records, majority of the products are not real animals (or animal products). As illustrated in Figure~\ref{fig:teaser}, there are postcards, plush toys, and decorative items that are returned for queries that search for wildlife. 
To filter out \emph{irrelevant} products, we use \textit{text-classification} and \textit{zero-shot-classification} using large-language models (LLMs). We can choose any textual attribute as input to the classifiers, in our dataset we used the name of the product.
The pipeline is flexible with respect to the model it uses to perform the task, e.g., we can choose any model available on HuggingFace~\cite{huggingface@github} for zero-shot or text classification.

For Zero-Shot Classification, we use a fine-tuned model of \textit{XLM-RoBERTa} \cite{conneau2019unsupervised}. The model \cite{xlm-roberta-large-xnli} is fine-tuned on a combination of data in 15 languages, but can also be effective in other languages since RoBERTa was trained in 100 different languages. 
We provide the model one hypothesis: \textit{`This product advertisement is about {}.'}, and the candidate labels (\textit{a real animal, a toy, a print of an animal, an object, a faux animal, an animal body part, a faux animal body part}). 
The output consists of the labels and their respective probabilities, which are added as attributes in our dataset. 

\myparagraph{Implementation Details} The pipeline runs on a Kubernetes cluster and is deployed with Docker images. The information extraction and model classification are executed as Kubernetes Jobs, sending data to S3-like object storage (\minio) as parquet files. We can access the data, both on Elasticsearch and \minio, from a JupyterHub also available on the cluster, and using DuckDB, we can execute SQL-style queries directly from the object store. This makes it possible to easily access and explore the data.

\section{The WildLife Ad Data Collection}
\label{sec:dataset}


\begin{figure}[t]
    \centering
    \begin{subfigure}[b]{0.49\textwidth}
        \centering
        \includegraphics[width=0.95\textwidth]{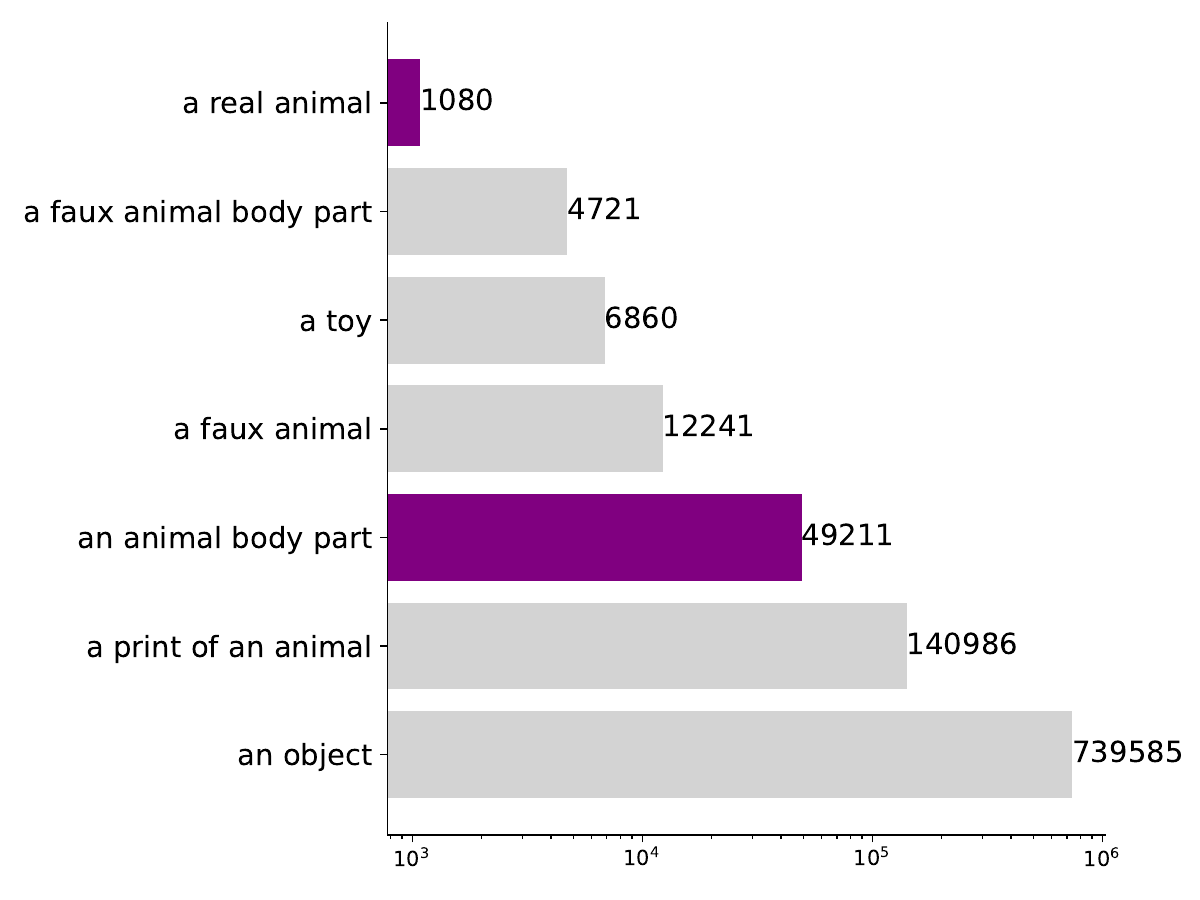}
        \caption{Distribution of zero-shot-classification labels from hypotheses: \textit{``This product advertisement is about:".}
        }
        \label{fig:distribution}
    \end{subfigure}
    \hfill
    \begin{subfigure}[b]{.49\textwidth}
        \centering
        \includegraphics[width=0.9\textwidth]{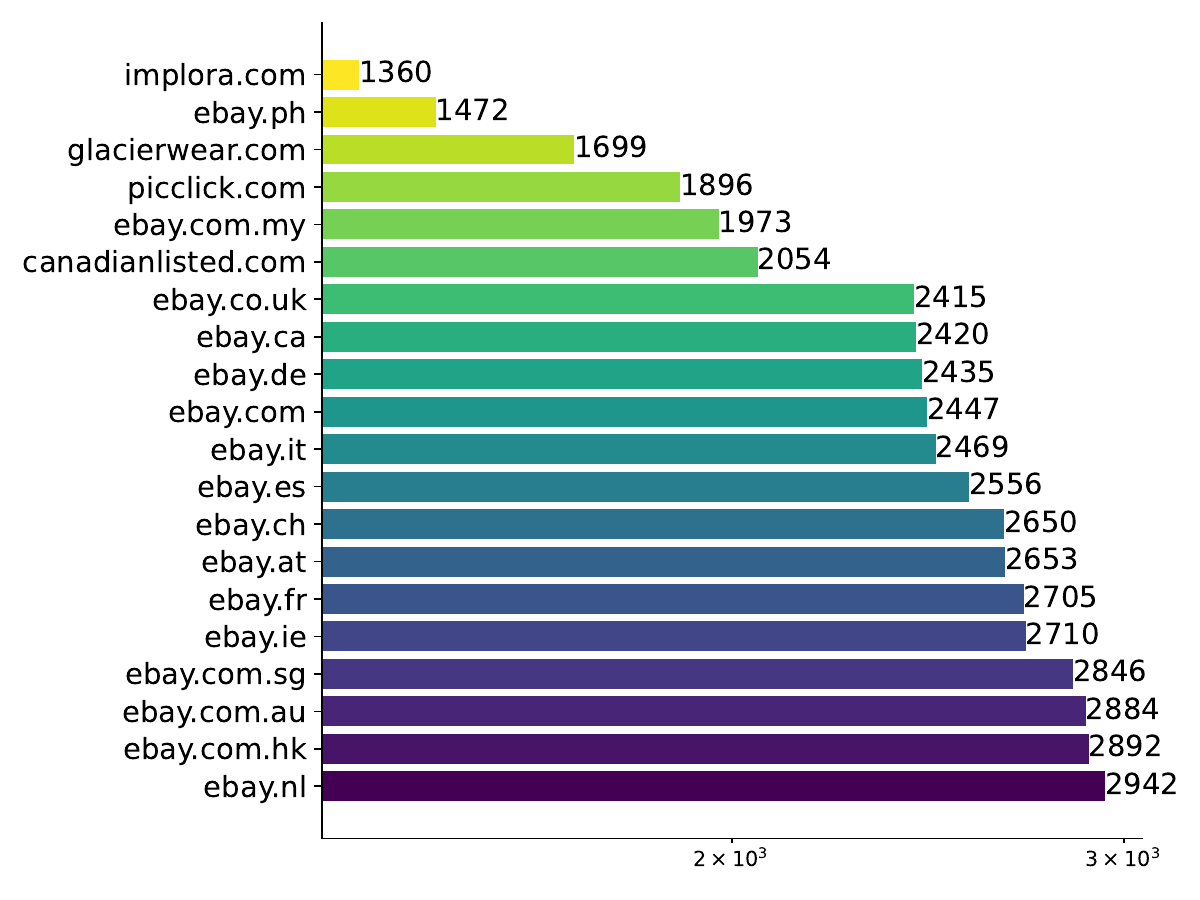}
        \caption{Top 20 domains where real animal and an animal body part classes are founded.
        \vspace{1em}
        }
        \label{fig:distribution domain}
    \end{subfigure}
    \caption{Distribution of zero-shot classes and domains.}
    \vspace{-.3cm}
\end{figure}

\myparagraph{Data Throughput}
To give an idea of the crawler throughput, we deployed a crawler on 2023-08-08.
As of 2023-09-05 (after 34 days), it had retrieved over 11 million pages.
The average time it takes to fetch a single page is 695.50 milliseconds (ms). This represents the typical response time for page retrieval. Due to politeness constraints, we avoid overloading the web servers where the pages reside, this way we limit the number of requests to the same server.
Running the actual pipeline takes longer: the current implementation processes approximately 145,000 pages per day.

\myparagraph{Data Overview}
We tested our pipeline using 954,684 pages and derived the dataset described in
\Cref{table: dataset-description}.
Each record corresponds to a web page and includes the URL, domain,  as well as the time when the page was downloaded. 
As expected, not all pages contain all attributes.
For instance, the price is available for 682,652 pages. 
Some attributes are rarer -- seller information was extracted for only 8,910 pages and location for 25,150.

\autoref{fig:distribution} shows the distribution of the zero-shot classes in the dataset. The classifier identified 1,080 products as being ``real animals" and 49,211 as being ``an animal body part". While the zero-shot classifier is not fool-proof, the low percentage of (potentially) relevant products underscores the importance of having an automated data collection and processing pipeline. 

The ability to scour a very large number of pages and sites opens the opportunity to obtain data at a scale not previously possible. For example, our collection includes sites in different countries.  \autoref{fig:distribution domain}  shows the top 20 domains in which the classes, ``a real animal" and ``an animal body part", are found. We can notice that from these 20 domains, 8 are from domains of non-English speaking countries, such as the Netherlands and Hong Kong.

\section{Discussion}

The pipeline we designed and implemented as a first step towards enabling large-scale data collection for wildlife products sold on the Web. Using this pipeline, it is possible to collect datasets that cover a wide range of data sources, in different countries, and that include a large set of species. We have designed the system to be flexible, i.e., it can be configured for specific tasks such as collections that focus on a specific species or sites; and by making it open source, we hope that the community will be able to collectively create and share datasets that can provide insights into wildlife trafficking.

An important goal of our design was to make the pipeline extensible. In this paper, we describe our current implementation and specific choices we have made for the different components. However, it is possible to replace these components. For example, alternative classifiers could be applied for data filtering, and different scraper mechanisms can be used for extraction.

There are a number of directions we intend to pursue in future work to improve our pipeline. The use of zero-shot models for data filtering is promising, but there is room for improvement. We are exploring the use of fine-tuned models such as DistilBert~\cite{sanh2019distilbert} as well as multi-modal models (which use both text and images)~\cite{yin2023survey} to improve the classification accuracy, in particular, to distinguish animal products from toys, prints,  clothes, etc. 

While rule-based systems such as MLScraper greatly simplify the task of extracting structured information from unstructured web pages, they are brittle and can fail in practice.
We would like to investigate the integration of deep-learning-based techniques (e.g., \cite{wang2022webformer, wang2022smartave}) to
create extractors that are robust and able to perform extraction from diverse websites, without the need for site-specific training.

\begin{acknowledgments}
  This work was funded by the NSF Disrupting Operations of Illicit Supply Networks (D-ISN) program.  
\end{acknowledgments}

\bibliography{paper}

\appendix

\end{document}